# Effect of Electron-Phonon Coupling on Thermal Transport across Metal-Nonmetal Interface – A Second Look

Xufei Wu[1], Tengfei Luo[1,2,*]

*1. Aerospace and Mechanical Engineering, University of Notre Dame*
*2. Center for Sustainable Energy at Notre Dame, University of Notre Dame*
*\* address correspondence to: tluo@nd.edu*

PACS: 73.40.Ns – Metal-Nonmetal Contacts

PACS: 73.40.−c – Electronic Transport in Interface Structures

PACS: 68.35.Md – Surface Thermodynamics, Surface Energies

PACS: 05.70.Np – Interface and Surface Thermodynamics

**Abstract:**

The effect of electron-phonon (e-ph) coupling on thermal transport across metal-nonmetal interfaces is yet to be completely understood. In this paper, we use a series of molecular dynamics (MD) simulations with e-ph coupling effect included by Langevin dynamics to calculate the thermal conductance at a model metal-nonmetal interface. It is found that while e-ph coupling can present additional thermal resistance on top of the phonon-phonon thermal resistance, it can also make the phonon-phonon thermal conductance larger than the pure phonon transport case. This is because the e-ph interaction can disturb the phonon subsystem and enhance the energy communication between different phonon modes inside the metal. This facilitates redistributing phonon energy into modes that can more easily transfer energy across the interfaces. Compared to the pure phonon thermal conduction, the total thermal conductance with e-ph coupling effect can become either smaller or larger depending on the coupling factor. This result helps clarify the role of e-ph coupling in thermal transport across metal-nonmetal interface.

# I. INTRODUCTION:

Thermal transport across metal-nonmetal interfaces is of great importance for applications such as microelectronics, thermoelectrics and electro-optical devices.[1, 2] In metals, electrons and phonons can both transfer heat, while in nonmetals thermal transport is exclusively dominated by phonons. Since phonon is the common heat carriers in metals and nonmetals, it is generally accepted that thermal transport across their interfaces is mediated by phonon-phonon coupling.[3, 4, 5] However, electron and phonon in metals are not isolated but interacting through electron-phonon (e-ph) coupling, enabling energy exchange between them. Whether such coupling is important for interfacial thermal transport is still under debate.

Majumdar and Reddy[3] used a two-temperature model to explain the role of e-ph coupling in metal as a thermal resistance between electron and phonon subsystems. This adds additional resistance to the intrinsic phonon-phonon interfacial resistance. However, Singh et al.[4] recently used the Bloch-Boltzmann-Peierls formula[6] and concluded that the e-ph coupling does not contribute significantly to the interfacial thermal conductance. One effect that has not been considered in these models is that the inelastic scattering between electron and phonon can perturb the phonon subsystem and can potentially results in a re-distribution of the phonon energy in the metal. For example, the usual e-ph scattering processes at room temperature involve one electron converting from one state to another with the absorption or emission of a phonon.[7] Higher order inelastic scatterings involving multiple electrons and phonons are also possible depending on the specific materials and temperature. These effects will change the populations of the phonons involved in the e-ph scattering processes and thus influence the scattering within the phonon subsystem.[8, 9, 10] Such an influence in the energy exchange among phonons can potentially lead to better phonon-phonon interfacial thermal transport.[11] As a result, e-ph coupling can have two competing effects:

it can add additional resistance on top of the resistance from interfacial phonon reflection, but it can also improve the phonon-phonon conductance across the interface due to its perturbation effect on the phonon subsystem.

The two-temperature model has been incorporated into molecular dynamics (MD) simulations using Langevin dynamics to simulate the effect of e-ph coupling on the phonon system (e.g., atomic trajectory).[12] Such a method should be able to capture the aforementioned effect of e-ph coupling on the phonon transport across interfaces. However, different conclusions were obtained in literature. Wang et al.[13] used the two-temperature MD to calculate the thermal conductance at a silicon-copper interface. It is observed that by including the e-ph effect, interfacial thermal conductance is significantly reduced. Jones et al.[14] and Nuo et al.[15] simulated the metal-nonmetal interface using a similar method. However, the e-ph coupling effect is found negligible for the interfacial thermal transport.

In this paper, we revisit the e-ph coupling effect on thermal transport across metal-nonmetal interfaces. From the two-temperature model, we first derive a general expression of total thermal conductance as a function of e-ph coupling factor, thermal conductivities of electrons and phonons, phonon-phonon conductance. We find that the Majumdar-Reddy serial resistance model is a special case of the derived general formula. By performing MD simulation with the two-temperature model and interpreting the data using the derived model, we find that e-ph coupling can perturb the phonon subsystem and increase the phonon-phonon thermal conductance. Compared to a pure phonon-phonon interface, the overall thermal conductance including electron effects can be either smaller or greater due to the competing effects from the e-ph coupling.

## II. THEORY:

Majumdar and Reddy[3] first introduced the two-temperature model to explain the role of e-ph coupling in thermal conductance of a metal-nonmetal interface. Their derived model reads:

$$h_{total} = \frac{h_{pp} \cdot \sqrt{g \cdot k_p}}{h_{pp} + \sqrt{g \cdot k_p}} \tag{1}$$

where $h_{pp}$ is the phonon-phonon thermal conductance at the interface, $g$ is the e-ph coupling factor and $k_p$ is the phonon thermal conductivity of metal. Such an expression was obtained by solving the following coupled equations for electron and phonon subsystems in the metal side, respectively:

$$k_e \frac{d^2 T_e(x)}{dx^2} - g \cdot (T_e(x) - T_p(x)) = 0 \tag{2a}$$

$$k_p \frac{d^2 T_p(x)}{dx^2} + g \cdot (T_e(x) - T_p(x)) = 0 \tag{2b}$$

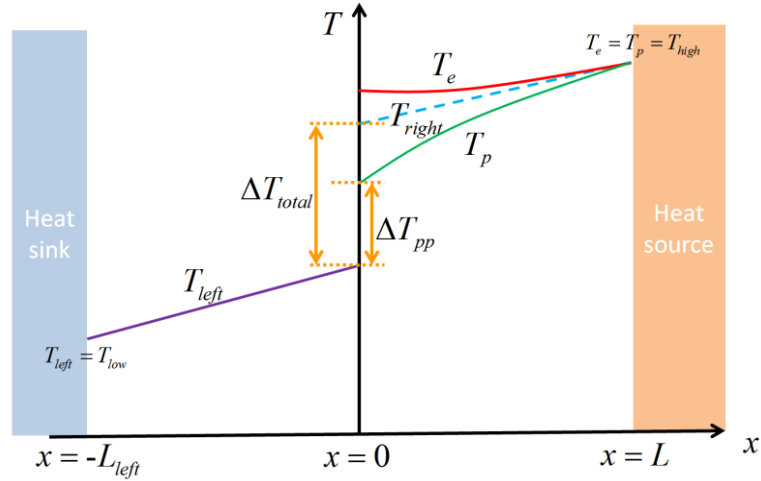

*Figure 1: Schematic plot describing two-temperature model for a metal-nonmetal interface.*

The solutions to these equations are in the following forms (see Fig. 1):

$$T_e(x) = A \cdot x + B + C \cdot \exp\left(\frac{x}{l}\right) + D \cdot \exp\left(-\frac{x}{l}\right) \tag{3a}$$

$$T_p(x) = A \cdot x + B - C \cdot \frac{k_e}{k_p} \cdot \exp\left(\frac{x}{l}\right) - D \cdot \frac{k_e}{k_p} \cdot \exp\left(-\frac{x}{l}\right) \tag{3b}$$

where $k_e$ is electronic thermal conductivity, and $A$, $B$, $C$ and $D$ are constants. $l = \sqrt{\frac{k_p \cdot k_e}{g \cdot (k_p + k_e)}}$ can be regarded as a characteristic length of e-ph coupling.

Below we show the derivation of the formulation of the interfacial thermal conductance at the metal-nonmetal interface. Fourier's law results in a linear profile of the total temperature of metal (right hand side) and the temperature of nonmetal (left hand side) at either side of the interface (Fig. 1):

$$T_{right}(x) = A \cdot x + B \tag{4a}$$

$$T_{left}(x) = E \cdot x + F \tag{4b}$$

where $A$, $B$, $E$ and $F$ are constants. The boundary conditions and the conditions at the interface are listed below and are indicated in Fig. 1. It is worth noting that we have assumed that the electrons do not transfer heat across the interface, which is consistent with the argument in the Majumdar-Reddy model[3].

$$T_p(L) = T_e(L) = T_{high} \tag{5a}$$

$$T_{left}(-L_{left}) = T_{low} \tag{5b}$$

$$\left.\frac{dT_e(x)}{dx}\right|_{x=0} = 0 \tag{5c}$$

$$-k_p \left.\frac{dT_p(x)}{dx}\right|_{x=0} = -k_{left}\frac{dT_{left}(x)}{dx} = h_{pp} \cdot \left(T_p(0) - T_{left}(0)\right) \tag{5d}$$

where $L$ is the thickness of the right hand side material (metal) and $L_{left}$ is that of the left hand side material (nonmetal). Equation (5d) also indicates how the phonon-phonon interfacial thermal conductance, $h_{pp}$, is defined. Then the total thermal conductance at interface is defined as the ratio of total heat flux to the temperature difference at interface (note here that $T_{right}$ refers the total temperature of the metal):

$$h_{total} = \frac{-k_{left}\frac{dT_{left}(x)}{dx}}{T_{right}(0) - T_{left}(0)} \qquad (6)$$

Combining Eqs. (4), (5) and (6), we can express the total thermal conductance, $h_{total}$, as a function of $k_e, k_p, g, h_{pp}$ and $L$, we can derive the total interfacial thermal conductance:

$$h_{total} = \frac{k_p \cdot h_{pp} \cdot (k_e + k_p) \cdot \cosh\left(\frac{L}{l}\right)}{h_{pp} \cdot k_e \cdot l \cdot \sinh\left(\frac{L}{l}\right) + k_p \cdot (k_e + k_p) \cdot \cosh\left(\frac{L}{l}\right)} \qquad (7)$$

It is also worth noting that when $k_e=0$ (*e.g.*, electron is not contributing to heat transfer), $h_{total}=h_{pp}$. If we further apply the condition that $k_e \gg k_p$ and at thick metal limit $L/l \gg 1$, the total thermal conductance will further deduce to the Majumdar-Reddy model (Eq. (1)).[3] It can be seen from Eq. (7) that the total thermal conductance will always be smaller than phonon-phonon thermal conductance. However, the phonon-phonon thermal conductance, $h_{pp}$, itself should be a function of coupling factor, $g$, according to the aforementioned reason: e-ph interaction can perturb the phonon subsystem and redistribute energy among different phonons and thus facilitate interfacial phonon-phonon energy transport. Phenomenalogically, it is also easy to understand that the actual phonon temperature profile in metal and at the interface will be different under the same heat flux condition whether the e-ph interaction is present or not. As a result, due to this effect, it is possible that the total thermal conductance can be larger than the intrinsic phonon-phonon conductance. When the Majumdar-Reddy's model is used, such an effect was usually ignored, and $h_{pp}$ was usually obtained from mismatch models.[16, 17] These models are developed for phonon-phonon interfacial thermal conductance without the consideration of e-ph coupling effects.

In the rest of this paper, we show through MD simulations that $h_{pp}$ will increase as we increase the coupling factor $g$. Compared to the pure phonon heat transfer case, we find that the total thermal conductance including e-ph coupling effect can indeed become either smaller or larger, depending on the e-ph coupling factor $g$.

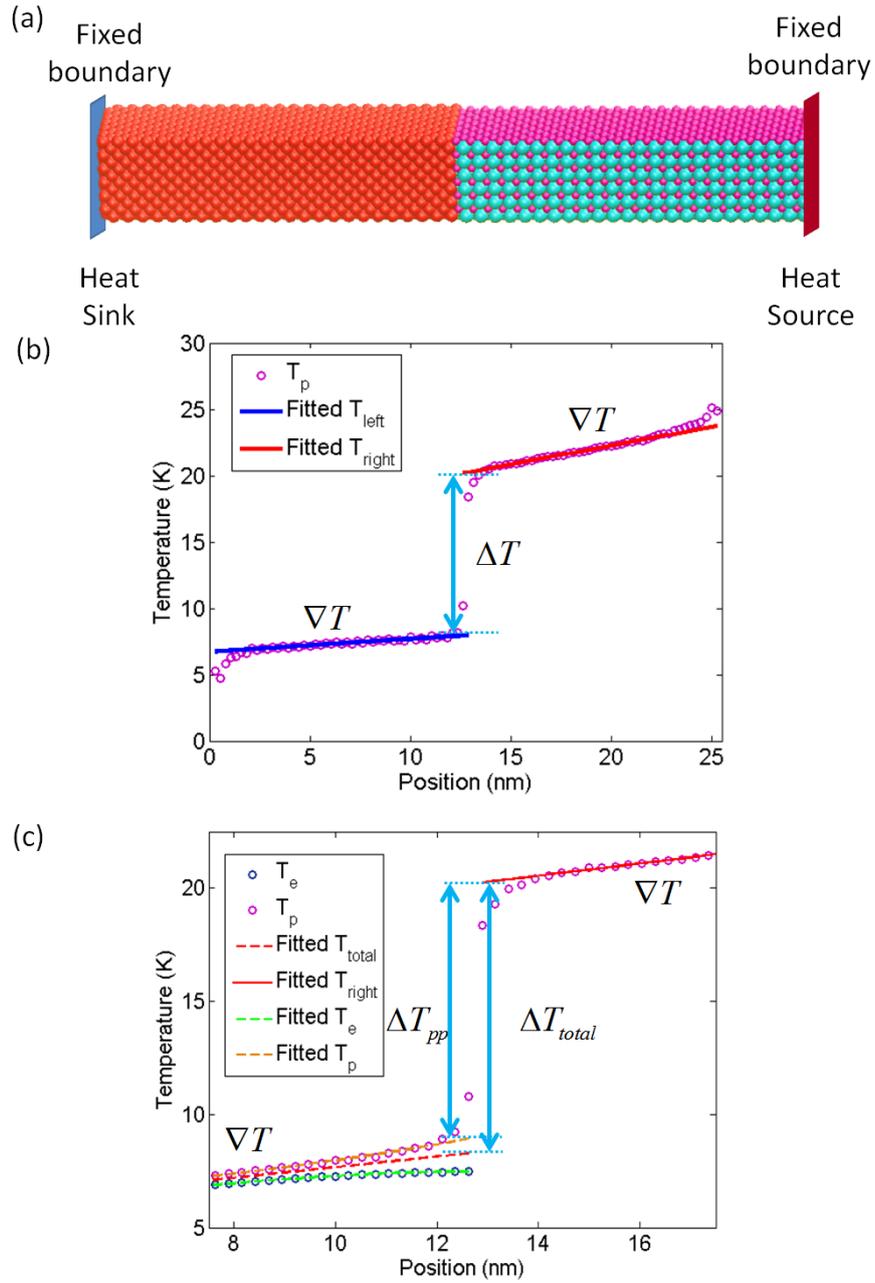

*Figure 2.* (*a*) MD simulation setup. (*b*) Steady state temperature profiles from an NEMD simulation without electron and (*c*) one with e-ph interaction. (Note: the x-axis in panel (*c*) is zoomed in for clarity. The full range plot is included in the supplemental information[18])

## III. RESULTS AND DISCUSSIONS:

MD simulations[19] are performed on a model bi-material junction as shown in Fig. 2a, and Langevin dynamics[20] is included for the atomic motion with the friction force and the stochastic force representing the effects of energy communication between electron and phonon subsystems. The detail of the simulation can be found in the supplemental information.[18]

We first study a case without e-ph interaction to obtain the intrinsic phonon-phonon thermal conductance, $h_{pp\_intrinsic}$. The total thermal conductance is calculated as $h_{pp\_intrinsic} = J/\Delta T$, where $J$ is heat flux crossing the interface which can be obtained by calculating the energies added to or taken out from the heat baths, and $\Delta T$ is obtained by extrapolating the linear portion of the temperature profiles in the materials and take the difference at the interface (Fig. 2b). We then performed MD simulations with e-ph interactions. Different e-ph coupling factors, $g$, and electronic thermal conductivity, $k_e$, are used. In these simulations, we strive to obtain two thermal conductances: $h_{total}$ and $h_{pp}$. Extracting $h_{total}$ is similar to how we extract $h_{pp\_instrinic}$. The only difference is that for the metal side, the linear extrapolation was done on the total temperature profile. However, the same method is not applicable to extracting $h_{pp}$ since the phonon temperature is not linear due to the e-ph coupling (Eq. (3b)). To obtain $h_{pp}$, we utilize Eq. (3) to fit the electron and phonon temperature profiles away from the interface. Fitting temperature profiles away from the interface is necessary since the nonlinearity of the phonon temperature near the interface is not due to e-ph coupling but rather the fact that atoms near the interface has difference vibrational feature from those inside the material and thus present additional resistance. This is because atoms near the interface will experience different force constants from those inside the material. Our theoretical model does not consider these effects. However, away from the interface, our model is valid. The fitted curve is extended to the interface and $h_{pp}$ is obtained utilizing the temperature difference of extrapolated phonon temperature profiles at the interface (see Fig. 2c). Using such a treatment, we have effective lumped the above-mentioned temperature bending effect into the interfacial thermal conductance data.

The interfacial thermal conductance as a function of electron-phonon coupling factor, $g$, and the electron thermal conductivity, $k_e$, are presented in Fig. 3. Electron thermal conductivity values chosen are 0.1, 1 and 10 W/mK. These values are respectively much smaller than, on the same order of magnitude as, and much larger than the phonon thermal conductivity of the lattice (~1W/mK),[11] roughly covering the range of highly doped semiconductors,[7] metallic carbon nanotube[21] to common noble metals.[13] For the e-ph coupling factor $g$, we choose values from $0.1 \times 10^{17}$ to $5 \times 10^{17}$ W/m$^3$K covers many common materials[22] (e.g., Ag: $0.2 \times 10^{17}$; Al: $2.45 \times 10^{17}$; Pt: $5 \times 10^{17}$ W/m$^3$K). It should be pointed out that the referred values are room temperature coupling factors. However, since MD simulations are classical, the simulations in this study, even though performed at a mean temperature of 15K, actually correspond to cases well above the Debye temperature. For most metals, the room temperature approaches this limit since they usually have relatively low Debye temperatures (e.g., Ag: 215K; Al: 428K; Pt: 240K). As a result, the use of room temperature coupling factor should still be relevant. It is also worth noting that there are some $h_{pp}$ points missing for the $k_e$=0.1 W/mK cases. It is because for these cases (e.g., small $k_e$ and large $g$), the fittings were not successful. Nevertheless, the available data are sufficient to support our following argument.

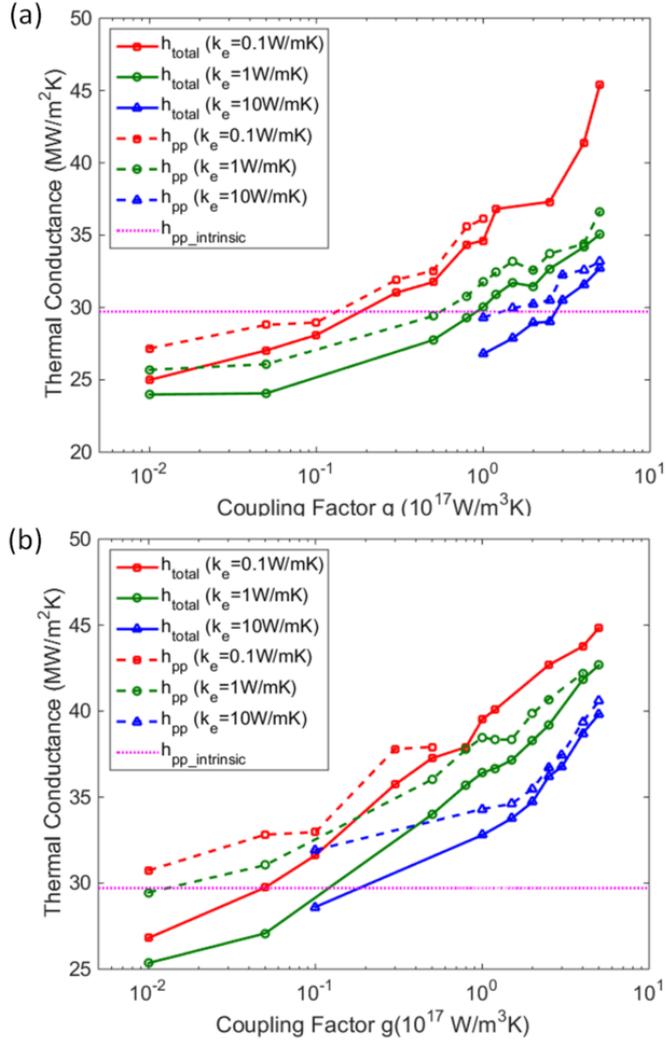

*Figure 3.* Panels (a) and (b) represent MD data from adding electron effects to the left material (monatomic lattic) and the right material (diatomic lattice), respectively. Solid and dashed curves represent the total thermal conductance and phonon-phonon thermal conductance, respectively. Red□, green ○ and blue △ represent the $k_e$ = 0.1, 1, and 10W/mK cases, respectively. The intrinsic phonon-phonon conductance, $h_{pp\_intrinsic}$ = 29.71 MW/m²K, is shown as a dashed horizontal line.

Figure 3a presents data for cases with the e-ph interaction applied to the monatomic lattice on the left hand side of the interface. These are common cases since most metals are monatomic. As seen from Fig. 3a, the total thermal conductance, $h_{total}$, increases monotonically when $g$ increases. Compared to the intrinsic phonon-phonon conductance ($h_{pp\_intrinsic}$, dashed horizontal line), $h_{total}$ can be either smaller or larger, depending on the electron thermal conductivity and coupling factor. We attribute such an observation to the competing effects e-ph coupling has on the interfacial thermal conductance, i.e., it adds an additional thermal resistance serially to $h_{pp}$ but can also enhance $h_{pp}$ itself due to its perturbation effect on the phonon system. This is proven in Fig. 3 which shows that $h_{pp}$ increases monotonically with $g$. It is also seen that $h_{total}$ is smaller than $h_{pp}$ in all cases. This agrees with Eq. (7) which indicates that e-ph coupling always add additional resistance to $h_{pp}$. Another observation worth mentioning is that the difference between $h_{pp}$ and $h_{total}$ becomes smaller as $g$ becomes larger. This is because larger $g$ facilitates the energy communication between electrons and phonons, and thus the resistance due to their coupling effects becomes smaller.

We have also simulated cases with the e-ph interaction applied to the diatomic lattice on the right hand side of the interface. Such simulations are also of practical relevance since materials like NbSe$_2$ and metallic carbon nanotube[21] are multi-atomic metals, and highly doped semiconductors can also have non-negligible electron contribution in heat transfer.[7] As shown in Fig. 3b, the trend of $h_{total}$ and $h_{pp}$ from these cases are the same as those in the previous cases shown in Fig. 3a. These cases also help prove that the perturbation due to e-ph coupling can enhance the energy communication between different phonon modes in metal. Such an effect was visualized by calculating the spectral temperatures of the optical and acoustic phonons in the diatomic lattice. The spectral temperature can be utilized to characterize how equilibrated different phonons are in terms of energy communication:[23, 24, 25] the closer the spectral temperatures are, the better the energy equilibration. The details of the calculation can be found in Ref. 9. It is found from Fig. 4 that larger e-ph coupling factor will increase the energy interaction between optical and acoustic phonons as indicated by the closer spectral temperatures of these two groups of modes. We have previously shown that such an enhanced internal energy communication will increase phonon thermal conductance at interface.[11]

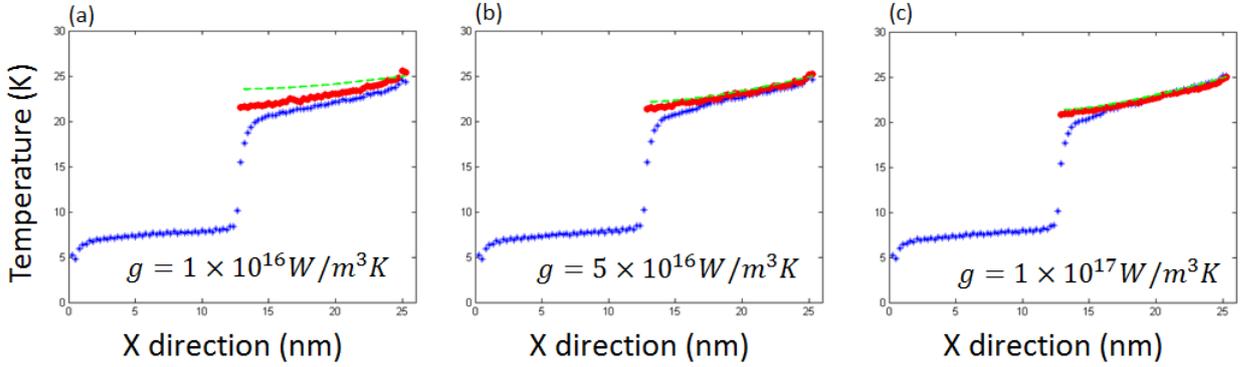

*Figure 4.* Steady state temperature profiles. For the diatomic lattice, the green dashed curve refers to electron temperature, the red curve refers to the spectral temperature of optical mode and the blue curve refers to that of acoustic mode. Cases with different e-ph coupling factors are shown. For all cases, $k_e$=1 W/mK.

Comparing Fig. 3a and 3b, we found that for the cases in Fig. 3b $h_{total}$'s are rarely smaller than $h_{pp\_intrinsic}$. The only difference between these two sets of cases is that in Fig. 3b the e-ph coupling effect is applied to the diatomic lattice. According to our previous study,[11] the interactions between optical and acoustic phonons can greatly influence the acoustic phonon transport across the interface. When e-ph coupling effect is applied to the diatomic lattice, such interactions can be enhanced. When e-ph coupling effect is applied to the monatomic lattice, which only has acoustic phonons, the interactions within acoustic modes are enhanced. While energy exchange within acoustic phonons can also help redistributing energy to improve thermal transport across the interface, it is possible that the energy exchange between acoustic and optical phonons has a stronger effect on the interfacial thermal conductance. This could be the reason why we see a more pronounced enhancement effect of e-ph coupling on $h_{total}$ in Fig. 3b.

It is also notable that for a given e-ph coupling factor, $g$, the interfacial thermal conductance will increase as electron thermal conductivity, $k_e$, decreases. When $k_e$ is low (0.1 W/mK, red curves, Fig. 3), the dominant heat carriers are phonons. In this case, the perturbation effect of the e-ph coupling is more dominant. Such a perturbation to the phononic system will enhance the phonon-phonon interfacial thermal conductance. When the coupling factor is larger, such perturbation effect is stronger, and thus we see an increase in the phonon-phonon thermal conductance (Fig. 3).

On the other hand, when electron thermal conductivity is high ($k_e$=10 W/mK, green curves, Fig. 3), the electron will dominate the heat transfer. In this case, electrons need to transfer energy to phonons which then carry energy across the interface. The resistance role of the e-ph coupling thus dominates its effects on the interfacial thermal transport process. In these cases, the electron thermal conductivity is much larger than the phonon counterpart, and common metals like copper, aluminum and gold fall into this category. In this scenario, the Majumdar-Reddy picture[3] should be applicable.

For the previously mentioned two-temperature MD works,[13, 14, 15] we can qualitatively explain the discrepancy among different simulations based on the conclusion obtained from the present study. For Wang's work,[13] copper was used, while aluminum was used in the other two studies.[14, 15] Copper has a much smaller e-ph coupling factor ($g$=0.5×$10^{17}$ W/m$^3$K) and a larger electron thermal conductivity than those of aluminum ($g$=2.4×$10^{17}$ W/m$^3$K),[22] respectively. According to Fig. 3a, e-ph coupling in copper is expected to present much larger thermal resistance than aluminum, which was found from these simulations.[13, 14, 15]

## IV. CONCLUSION:

In this paper, we found that the phonon-phonon thermal conductance at metal-nonmetal interfaces increases as the e-ph coupling factor in the metal side increases. This is because energy communications between different phonon modes in metal is strengthened by e-ph interactions which facilities energy redistribution into modes that can more easily transfer energy through the interfaces. It is found that the serial thermal resistance modal proposed by Majumdar and Reddy is applicable under high electron thermal conductivity and small e-ph coupling situations. The total thermal conductance may be smaller or larger than the intrinsic phonon-phonon conductance depending on the actual e-ph coupling factor and thermal conductivity of electrons in real materials.

**ACKNOWLEDGMENT:**

This research was supported in part by the Notre Dame Center for Research Computing and NSF through XSEDE resources provided by SDSC Trestles and TACC Stampede under grant number TG-CTS100078. T.L. thanks the support by the Semiconductor Research Corporation (contract number: 2013-MA-2383) and the startup fund from the University of Notre Dame. We thank Dr. Bo Qiu for useful discussions.

# Supplemental Information

# Effect of Electron-Phonon Coupling on Thermal Transport across Metal-Nonmetal Interface – A Second Look

Xufei Wu, Tengfei Luo

**1. Interatomic Potential using Force Constants**

In general, the potential energy (V) and force ($\vec{F}$) can be expanded in Taylor Series in terms of the atomic displacement ($\vec{u}$) from their equilibrium positions: [1]

$$V = V_0 + \sum_i \alpha_i u_i + \frac{1}{2!}\sum_{i,j}\beta_{i,j}u_i u_j + \frac{1}{3!}\sum_{i,j,k}\gamma_{i,j,k}u_i u_j u_k + ... \quad (s1)$$

$$\vec{F}_i = -\frac{\partial V}{\partial \vec{u}_i} = -\alpha_i - \sum_j \beta_{i,j}\vec{u}_j - \frac{1}{2!}\sum_{j,k}\gamma_{i,j,k}\vec{u}_j \vec{u}_k - ... \quad (s2)$$

Where $\alpha, \beta, \gamma$ are the first, second and third order force constants, *i, j, k* denote the indices of atoms. The first order force constants $\alpha$ are usually zero as atoms should have no force at equilibrium positions. The second order force constants $\beta$ are also called the harmonic force constants, which determine the phonons vibrational properties, such as phonon dispersion relations and phonon density of states. The third order force constants $\gamma$, the anharmonic force constants, can influence the three phonon scattering processes. The force constants are limited to third order here as we do not expect effects of higher order terms to be significant on thermal transport. The specific values of the force constants are extracted from Lennard-Jones Argon model. Modifications to the force constants are made to align the acoustic peak frequency for materials at the two sides of the interface. A list of force constants are shown in Ref. [3] Table I.

**2. Langevin Dynamics with E-Ph coupling**

To include e-ph coupling effect, Langevin dynamics is used for the atomic motion with the friction force and the stochastic force representing the effects of energy communication between electron and phonon subsystems.[2] The Langevin scheme is applied to every atom individually. The total force is expressed as:

$$\vec{F}_i = -\frac{\partial V}{\partial \vec{u}_i} - \mu_i \frac{d\vec{u}_i}{dt} + \tilde{F}_i(t) \qquad (s3)$$

where the damping constant is related to the e-ph coupling factor as:

$$\mu_i = \frac{gm_i}{3nk_B} \qquad (s4)$$

where $n$ is the number density of atoms, $m$ is the atomic mass, and $k_B$ is Boltzmann constant. From the fluctuation-dissipation theorem, the stochastic force $\tilde{F}(t)$, is a Gaussian random variable with mean and variance as:

$$\langle \tilde{F}_i(t) \rangle = 0 \text{ and } \langle \tilde{F}_i(t)\tilde{F}_j(t+\Delta t) \rangle = 2k_B T_e \mu_i \delta_{ij} \delta(\Delta t) \qquad (s5)$$

The electron temperature, $T_e$, is predicted using heat conduction equation considering the energy exchange between electron and phonon subsystems:

$$C_e \frac{dT_e(x,t)}{dt} + g\left(T_e(x,t) - T_p(x,t)\right) = k_e \frac{d^2 T_e(x,t)}{dx^2} \qquad (s6)$$

where $C_e$ is the specific heat capacity of electron, $k_e$ is the thermal conductivity of electrons and $T_p$ is the local phonon temperature. From reference[4], we assume that $C_e$=100 J/m$^3$K$^2$×Te, which is proportional to the temperature of electron. It is noted that, for common metals, the ratio of $C_e$ and $T_e$ is on the order of 100 J/m$^3$K$^2$, and thus it is used in this work. It is worth noting that since all the analyses are performed at the steady state, the choice of $C_e$ does not influence our result.

### 3. Simulation Details

MD simulations are performed on a model bi-material junction, which consists of a monatomic and a diatomic lattice with the same lattice constant and interatomic potential. Both lattices have face-centered cubic structure. The two kinds of atom of the diatomic lattice are assigned to different masses to create a phonon band gap between acoustic phonons and optical phonons. Masses of atoms are chosen so that the cutoff frequencies of acoustic phonons of both materials are close to each other. For the monoatomic side, the mass is 99.87, and those on the diatomic side are 23.97 and 63.92. The potential energy and force of atoms are described by the force constant method in Section 1.

In the simulations, the temperature of electron is updated every MD step by solving discretized Eq. (s6) using a finite difference method. The temperature of phonons at each time step is averaged layer by layer with each layer containing 50 atoms. The cross section of the simulation supercell has a size of 2.63×2.63 nm$^2$, and the lengths are chosen to meet the $L/\lambda \gg 1$ limit. We use periodical boundary conditions in the cross-sectional directions and fixed boundary condition in the heat transfer direction. The system temperature is maintained at 5 and 25 K for each end, respectively, using velocity scaling. Time step is set to 1 fs.

**4. Cross-sectional Size Effects on Interfacial Thermal Conductance**

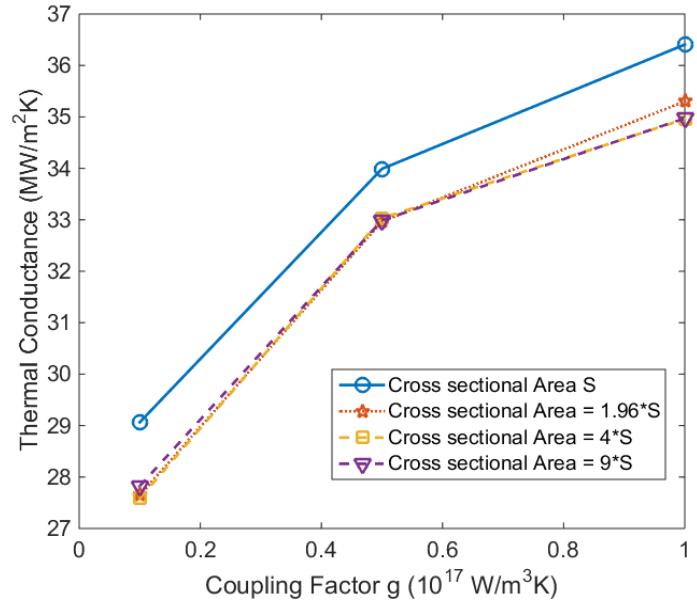

*Figure S1. Total interfacial thermal conductance $h_{total}$ as a function of coupling factor g, for systems with different cross-sectional areas. Here $k_e$ =1 W/mK. The blue curve is the original curve in Fig. 3(b) in the main text. Here, S refers to the original size.*

MD simulations for systems with different cross sectional areas (*3.68 x 3.68, 5.26 x 5.26, 7.89 x 789 nm$^2$, see the red, orange* and purple curve in Fig. S1) have been studied. It is found that the trend of interfacial thermal conductance change as a function coupling factor is the same for all cross-sectional sizes. Therefore, the size effects will not change our conclusion in paper.

## 5. Length Effects on Interfacial Thermal Conductance

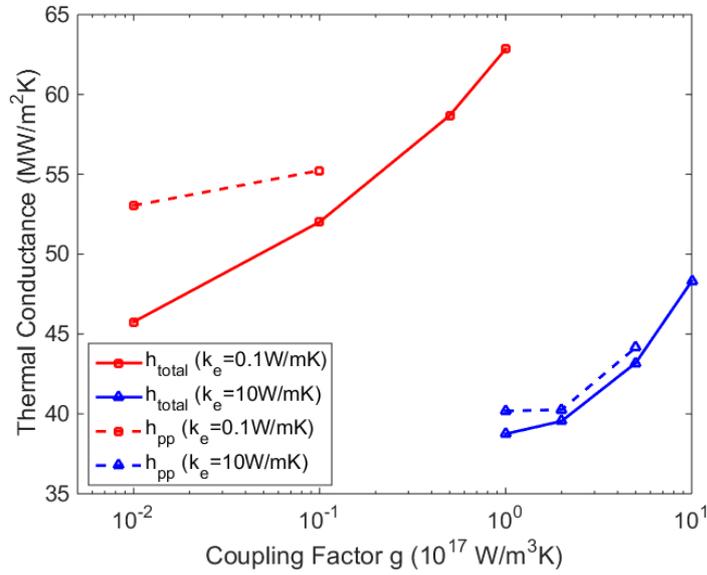

*Figure S2. Interfacial thermal conductance as a function of coupling factor g. The length of the system is doubled compared to the system studied in main text.*

In the length dependent study, we found that the thermal conductance increased as we enlarge the length of the system. It can be easily explained as that there are more low frequency modes in systems with longer length, which usually have higher transmissivity across the interface. However, the trend of $h_{pp}$ and $h_{total}$ is the same as that observed in the smaller system presented in the main text (Fig. S2). Since this is a model study which does not correspond to any realistic system, we utilized the data obtained from the small system to study the physics.

## 6. Temperature profile with electron phonon interaction

Figure s3 here shows the original temperature profile with electron phonon interaction with full range (from 0 to 25.25 nm). The fitted electron and phonon temperature overlaps with each other in this figure, which makes it hard to read. Therefore, we adopt a zoom-in figure in paper (Figure. 2(c)) for clarity.

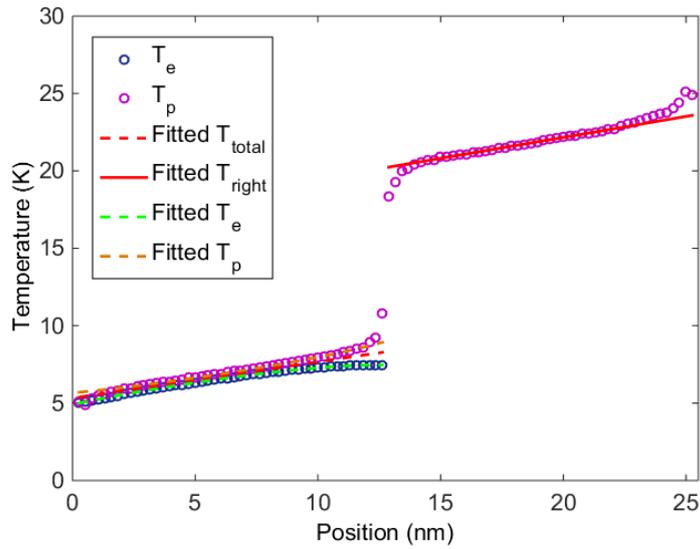

*Figure S3. Steady state temperature profiles from an NEMD simulation with e-ph interaction. The x-axis is in full range.*